# Theoretical Modeling for the Interaction of Tin alloying with N-Type Doping and Tensile Strain for GeSn Lasers

David S. Sukhdeo[1], Krishna C. Saraswat[1], Birendra (Raj) Dutt[2,3], and Donguk Nam[*,4]

[1]*Department of Electrical Engineering, Stanford University, Stanford, CA 94305, USA*

[2]*APIC Corporation, Culver City, CA 90230, USA*

[3]*PhotonIC Corporation, Culver City, CA 90230, USA*

[4]*Department of Electronics Engineering, Inha University, Incheon 402-751, South Korea*

*Email: dwnam@inha.ac.kr

**Abstract:** We investigate the interaction of tin alloying with tensile strain and n-type doping for improving the performance of a Ge-based laser for on-chip optical interconnects. Using a modified tight-binding formalism that incorporates the effect of tin alloying on conduction band changes, we calculate how threshold current density and slope efficiency are affected by tin alloying in the presence of tensile strain and n-type doping. Our results show that while there exists a negative interaction between tin alloying and n-type doping, tensile strain can be effectively combined with tin alloying to dramatically improve the Ge gain medium in terms of both reducing the threshold and increasing the expected slope efficiency. Through quantitative modeling we find the best design to include large amounts of both tin alloying and tensile strain but only moderate amounts of n-type doping if researchers seek to achieve the best possible performance in a Ge-based laser.

## *Introduction*

Optical interconnects have garnered much attention in recent years as a potential solution to the problems that plague modern electrical interconnects [1], [2]. The most significant challenge in realizing silicon-compatible optical interconnects is the challenge of creating a CMOS-compatible light source [2]–[4]. Light sources based on Group-IV materials have been researched extensively including silicon Raman lasers [5], [6] and Ge lasers [7]–[10]. Although electrically-pumped Ge lasers with heavy n-type doping have been demonstrated [7], [8], the observed thresholds were extremely high because of the challenges posed by Ge's indirect bandstructure. N-type doping helped somewhat with this problem but was on its own insufficient [11]. Researchers have therefore proposed employing band engineering to transform Ge into a direct bandgap material that is better suitable for lasing [11]–[17]. The two leading forms of band engineering for achieving direct bandgap Ge are tin alloying [12]–[14] and tensile strain [11], [15], [16], both of which lower the direct Γ conduction valley faster than indirect L valleys [17]. However, strain and tin alloying techniques have been progressing separately from each other. For example, direct bandgap Ge has been demonstrated by using tin alloying alone [18], using uniaxial tensile strain alone [19], and researchers are close to achieving direct bandgap Ge using only biaxial tensile strain [20]. This notion of two separate research tracks also holds true for laser device realization. One track focuses on tensile strained germanium lasers [15], [21] and another on GeSn lasing [18], yet so far neither of these efforts has delivered a practical device. Meanwhile, it has not been explored even theoretically whether or not the tin alloying and tensile strain can be combined in a useful manner for creating a practical Ge laser, and the answer to this question is not obvious. Tensile strain showed a negative interaction with n-type doping, for instance, such that doping offered comparatively smaller benefits in the presence of tensile strain [11]. In this paper, we are therefore interested to know how tin alloying interacts with tensile strain as well as n-type doping in the context of Ge-based laser device and what this implies for the design of practical tensile strained germanium-tin (GeSn) lasers. Here we will show that a positive interaction exists between tensile strain and tin alloying such that tensile strained GeSn is indeed a promising material for practical laser devices.

## *Bandstructure modeling*

The effect of tin alloying on Ge's bandstructure can be modeled in a number of ways such as density functional theory [17], tight-binding [22] or empirical pseudo-potentials [12]; in all cases the virtual crystal approximation with disorder effects [12], [17], [22] can be used to account for the combination of two dissimilar elements. Similar options are available for computing Ge's bandstructure under strain [11], [17], [23], [24]. In this work we use tight-binding to compute the bandstructures of Ge over a mesh of k-points encompassing the entire first Brillouin Zone and then repeat this lengthy computation for a range of tensile strain values. Because repeating this intensive calculation over a matrix of possible strains and tin concentrations would be computationally infeasible, we assume that germanium tin changes the conduction band in qualitatively the same way as tensile strain, i.e. reducing the direct gap relative to the indirect gap, but without substantially altering the valence band, i.e. without the LH/HH splitting that occurs under biaxial and uniaxial tensile strain. Assuming that a direct gap occurs at either 6.55% tin [12] or 2.40% biaxial tensile strain (<100> orientation) [11], we take every 1% tin to equal 0.3664% biaxial strain for the purposes of the conduction band. We then combine the conduction bands from our tight-binding model for the appropriate strain level with the valence bands for unstrained germanium and then, following the example of the scissor operator [25] commonly used in density functional theory, we manually raise or lower the conduction bands such that the bandgap is forced to the correct value in accordance with Ref. [26]. This results in a bandgap dependence on tin concentration as shown in Fig. 1(a) which is in good agreement with accepted models and experimental results for tin concentrations smaller than about 10% [12], [26]–[28]. This approach of directly deriving GeSn bandstructures from the strained Ge bandstructures makes it computationally feasible to generate the bandstructures over a mesh comprising the entire first Brillouin zone for every possible combination of tensile strain and tin concentration. Thus, our highly efficient approach to computing bandstructures is what enables us to explore the device implications of combining strain and tin, in contrast with previous work on computing bandstructures of strained GeSn which limited the bandstructure analysis to a very small number of critical k-points [17]. Most importantly, we find that our computed band edges for relaxed GeSn, shown in Fig. 1(a), are in excellent agreement with accepted models and experiments [12], [17]. Our models predict that Ge becomes a direct bandgap semiconductor at either 6.55% tin, which is in good agreement with published values [12], [17], or at 2.40% biaxial tensile strain, which is the same value that we obtained from our tight-binding model in [11]. We also find that our computed bandstructures for Ge in the presence of both biaxial strain and tin alloying simultaneously are also in agreement with accepted values [17], confirming the suitability of our computer bandstructures for use in optical device modeling.

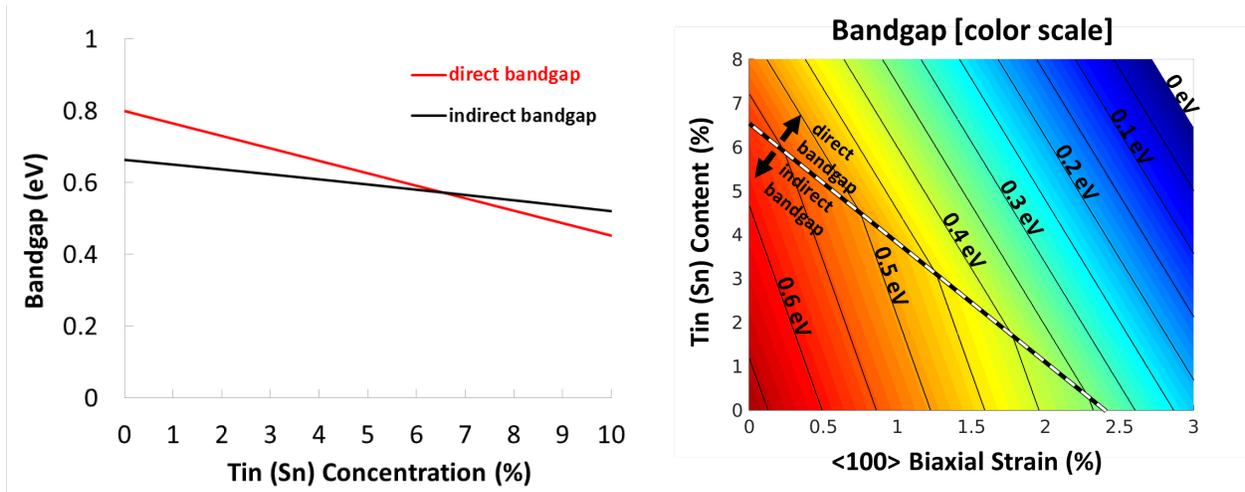

*Fig. 1.* *(a) GeSn's direct and indirect bandgap energies vs. tin concentration assuming zero strain. Crossover of the direct gap is visible at 6.55% tin. (b) Color mapping of the bandgap as a function of biaxial tensile strain and tin concentration.*

Having computed the bandstructure of strained GeSn, the next step is to model how tin alloying interacts with n-type doping and tensile strain to affect the performance of a germanium laser. As shown in Fig. 2(a), we find that band engineering through tin alloying up to 10% can reduce the threshold of a germanium-based laser by over two orders of magnitude. This assumes that the GeSn gain medium is completely unstrained and the results are in excellent agreement with prior GeSn laser modeling using empirical pseudopotential method [12], thus validating our approach for computing bandstructures in this work. Note that the lower left region of Fig. 2(a) is not available due to the cutoff of our simulation bounds which exclude thresholds beyond 1000 kA/cm$^2$. By re-plotting Fig. 2(a) for the threshold vs. tin content for various doping levels, we observe in Fig. 2(b) that n-type doping becomes quite harmful at higher tin concentrations. At 1% tin concentration, for example, increasing the doping from 1e18 cm$^{-3}$ to 1e20 cm$^{-3}$ can reduce the threshold more than one order of magnitude. However, at 9% tin concentration, increasing the doping from 1e18 cm$^{-3}$ to 1e20 cm$^{-3}$ increases the threshold more than one order of magnitude. This negative interaction between tin alloying and n-type doping means that even though tin alloying and n-type doping are each quite useful on their own, carelessly combining these two techniques may have disastrous results at the device level.

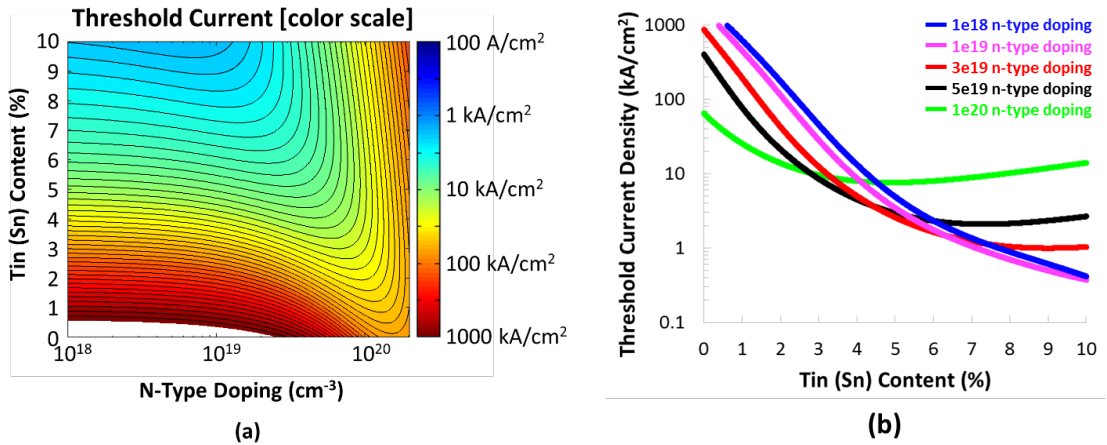

*Fig. 2.* *(a) Threshold current density of an unstrained GeSn laser (color scale) vs. tin content and n-type doping. (b) Threshold current density of a GeSn laser vs. tin content for different doping conditions. In all cases the GeSn thickness is assumed to be 300nm with a perfect double heterostructure, an optical cavity loss of zero, and a defect-limited minority carrier lifetime of 100ns. The blank region in the bottom left corner is due to the cutoff of the simulation bounds, i.e. thresholds greater than 1000 kA/cm$^3$.*

The next task is to investigate how tin alloying interacts with tensile strain with respect to Ge-based laser performance. Fig. 3(a) shows a 2D color mapping of how tensile strain and tin alloying work together to effectively reduce the lasing threshold. Compared to relaxed pure Ge (i.e. zero strain and zero tin concentration), a combination of 5% tin and 2% tensile strain can reduce the threshold by 4 orders of magnitude. Most importantly, we find from Fig. 3(a) that there is no negative interaction whatsoever between tin alloying and biaxial tensile strain, a fact which is shown more explicitly in Fig. 3(b). Even at 5% tin concentration, i.e. $Ge_{0.95}Sn_{0.5}$, increasing the biaxial tensile strain from 0% to 2% can reduce the threshold by almost 2 orders of magnitudes. We therefore conclude that tensile strain and tin alloying can indeed be effectively combined for a low threshold laser, i.e. a tensile strained GeSn laser.

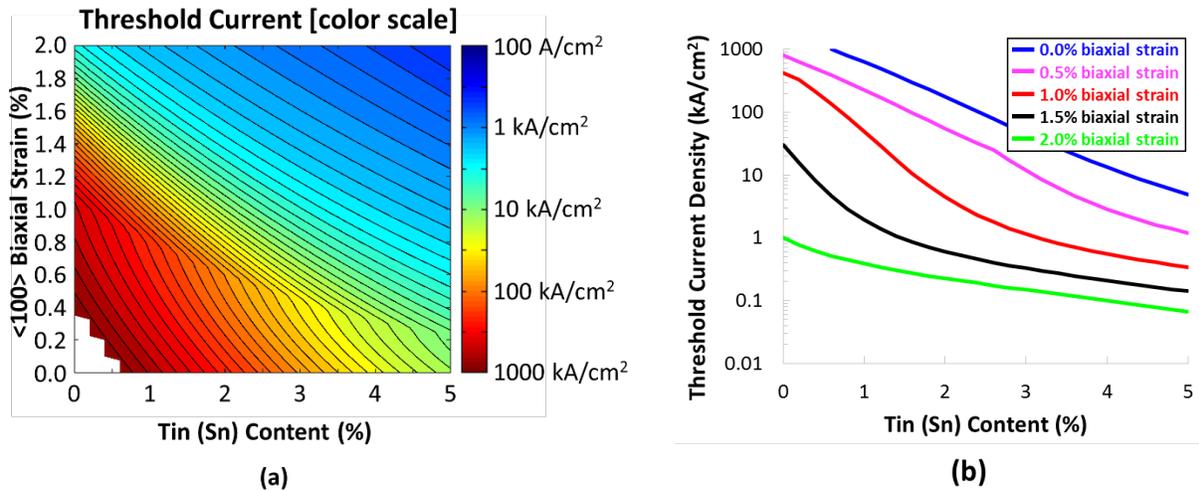

**Fig. 3.** *(a) Threshold current density of a double heterostructure GeSn laser (color scale) vs. <100> biaxial tensile strain and tin content. (b) Threshold current density of a GeSn laser vs. tin content for different amounts of biaxial tensile strain. In all cases the GeSn thickness is assumed to be 300nm with a perfect double heterostructure, an optical cavity loss of zero, and a defect-limited minority carrier lifetime of 100 ns. The blank region in the bottom left corner is due to the cutoff of the simulation bounds, i.e. thresholds greater than 1000 $kA/cm^3$.*

Having established that tensile strain and tin alloying can indeed be combined to achieve a low threshold Ge-based laser, the next question is whether or not such a combination would result in a useful slope efficiency. In Fig. 4 we calculate the slope efficiency vs. tin content and biaxial strain for different optical cavity losses. We need to consider different optical cavity losses because it very strongly affects the slope efficiency [11], [12]. Note that by "optical cavity loss" we refer only to useful out-coupling of light rather than parasitics such as scattering or optical losses in a metal electrode; such parasitics are completely ignored. From Fig. 4, we find that combining tensile strain with tin alloying does indeed result in a useful slope efficiency. Interestingly, for the case of 5% tin, we do observe an "ultimate limit" of about 1.5% biaxial strain whereupon increasing the strain further degrades the slope efficiency. This indicates that, when pushed to the limit, tin alloying offers greater performance enhancements to a Ge-based laser than biaxial strain, since no such limit was yet observed for tin alloying. Nevertheless, the overall finding is that comparatively large slope efficiencies are indeed possible when combining tensile strain with tin alloying, and thus a tensile stain GeSn laser is a very promising route to achieving both a low threshold and a reasonable slope efficiency.

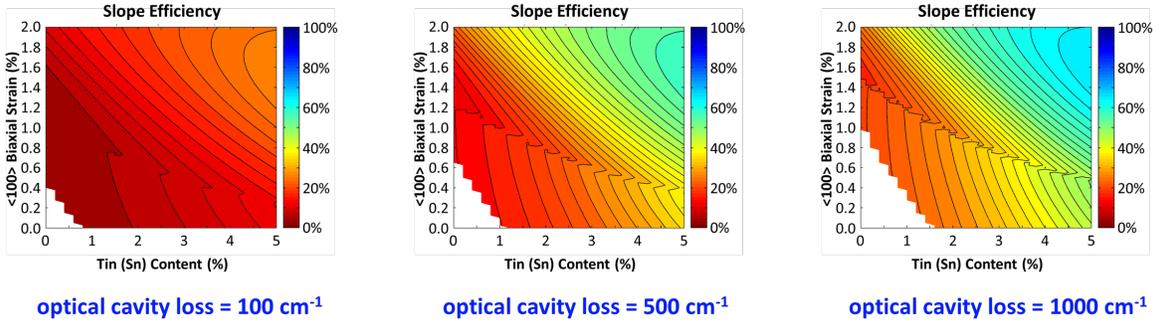

*Fig. 4.* Slope efficiency for a biaxially strained GeSn laser, shown for optical cavity losses of 100 $cm^{-1}$ (left), 500 $cm^{-1}$ (middle) and 1000 $cm^{-1}$ (right). In all cases, a double heterostructure design with a 300nm thick germanium tin ($Ge_xSn_{1-x}$) active region with $1\times10^{18}$ $cm^{-3}$ n-type doping and a defect-limited minority carrier lifetime of 100 ns is assumed.

## Conclusion

In summary, we have performed theoretical modeling of how tin alloying interacts with both n-type doping and tensile strain, i.e. for tensile strained n-doped GeSn lasers. It was found that there was a negative interaction between tin alloying and n-type doping: whereas n-type doping can help reduce the threshold when the tin content was low, the threshold was actually increased with n-type doping when the tin content was higher. In contrast, a combination of tin alloying and tensile strain was always effective at reducing the lasing threshold. Compared to unstrained Ge with no tin content, the threshold can be reduced by more than 4 orders of magnitudes if a combination of 2% tensile strain and 5% tin alloying is employed. We also found that tin alloying and tensile strain both help enhance the slope efficiency of GeSn lasers, and work very effectively together in this regard. These results suggest that a combination of heavy tin alloying and large tensile strain with only moderate n-type doping represents a very promising route to an efficient low-threshold Ge-based laser for on-chip optical interconnects.

*Acknowledgements*

This work was supported by the Office of Naval Research (grant N00421-03-9-0002) through APIC Corporation (Dr. Raj Dutt) and by a Stanford Graduate Fellowship. This work was also supported by an INHA UNIVERSITY Research Grant and by the Pioneer Research Center Program through the National Research Foundation of Korea funded by the Ministry of Science, ICT & Future Planning (2014M3C1A3052580). The authors thank Shashank Gupta of Stanford University and Boris M. Vulovic of APIC Corporation for helpful discussions. The authors also thank Ze Yuan of Stanford University for his help implementing the tight-binding code.